\documentclass[12pt,epsf]{article}
\usepackage{epsfig}

\newcommand{\onefigure}[2]{\begin{figure}[htbp]
\begin{center}\leavevmode\epsfbox{#1.eps}\end{center}\caption{#2\label{#1}}
\end{figure}}

\renewcommand{\thanks}[1]{\footnote{#1}} 

\newcommand{\be}{\begin{equation}}
\newcommand{\ee}{\end{equation}}
\newcommand{\bea}{\begin{eqnarray}}
\newcommand{\eea}{\end{eqnarray}}

\begin{document}

\begin{flushright}
SLAC-PUB-10571\\
26 July 2004\\
\end{flushright}

\bigskip\bigskip
\begin{center}
{\bf\large A Calculation of Cosmological Scale from Quantum
Coherence\footnote{\baselineskip=12pt Work supported by Department
of Energy contract DE--AC03--76SF00515.}}
\footnote{To be presented at The Twenty-Sixth Annual Meeting of the
ALTERNATIVE NATURAL PHILOSOPHY ASSOCIATION
Cambridge, England, 31 July - 5 August, 2004}
\end{center}

\begin{center}
James V. Lindesay\footnote{Permanent address, Physics Department, Howard University,
Washington, D.C. 20059}, jlslac@slac.stanford.edu\\
H. Pierre Noyes, noyes@slac.stanford.edu \\
Stanford Linear Accelerator Center MS 81,
Stanford University \\
2575 Sand Hill Road, Menlo Park CA 94025\\
\end{center}

\begin{center}
{\bf Abstract}
\end{center}
We use general arguments to examine the energy scales for which a
quantum coherent description of gravitating quantum energy units
is necessary.  The cosmological dark energy density is expected to
decouple from the Friedman-Lemaitre energy density when the
Friedman-Robertson-Walker scale expansion becomes sub-luminal at
$\dot{R}=c$, at which time the usual microscopic interactions of
relativistic quantum mechanics (QED, QCD, etc) open new degrees of
freedom.  We assume that these microscopic interactions cannot
signal with superluminal exchanges, only superluminal quantum
correlations. The expected gravitational vacuum energy density at
that scale would be expected to freeze out due to the loss of
gravitational coherence. We define the vacuum energy which
generates this cosmological constant to be that of a zero
temperature Bose condensate at this gravitational de-coherence
scale. We presume a universality throughout the universe in the
available degrees of freedom determined by fundamental constants
during its evolution. Examining the reverse evolution of the
universe from the present, long before reaching Planck scale
dynamics one expects major modifications from the de-coherent
thermal equations of state, suggesting that the pre-coherent phase
has global coherence properties. Since the arguments presented
involve primarily counting of degrees of freedom, we expect the
statistical equilibrium states of causally disconnected regions of
space to be independently identical.  Thus, there is no luminal ``horizon"
problem associated with the lack of causal influences between
spatially separated regions in this approach. The scale of the
amplitude of fluctuations produced during de-coherence of
cosmological vacuum energy are found to evolve to values
consistent with those observed in cosmic microwave background
radiation and galactic clustering.
\bigskip

\section{Introduction}
\indent

There is general (although not universal) agreement among physical
cosmologists that the current expansion phase in the evolution of
our universe can be extrapolated back toward an initial state of
compression so extreme that we can neither have direct laboratory
nor indirect (astronomical) observational evidence for the laws of
physics needed to continue that extrapolation.  Under these
circumstances, lacking a consensus "theory of everything", and in
particular a theory of "quantum gravity", we believe that the
prudent course is to rely as much as possible on general
principles rather than specific models.  This approach is adopted
in this paper.

We believe that the experimental evidence for currently accepted
theories of particle physics is relevant up to about 5 TeV---the
maximum energy or temperature we need consider in this paper. We
further assume that our current understanding of general
relativity as a gravitational theory is adequate over the same
range, and consequently that the cosmological Friedman-Lemaitre
(FL) (Hubble) dynamical equations are reliable guides once we have
reached the observational regime where the homogeneity and
isotropy assumptions on which those equations are based become
consistent with astronomical data to requisite accuracy. Although
the elementary particle theories usually employed in relativistic
quantum field theories have well defined transformation properties
in the flat Minkowski space of special relativity, we hold that
their fundamental principles still apply on coordinate backgrounds
with cosmological curvature. In fact there is direct experimental
evidence that quantum mechanics does apply in the background space
provided by the Schwarzschild metric of the earth thanks to the
beautiful experiments by Overhauser and
collaborators\cite{Over74,Over75}. These experiments show that the
interference of a single neutron with itself changes as expected
when the plane of the two interfering paths is rotated from being
parallel to being perpendicular to the ``gravitational field" of
the earth.

Since quantum objects have been shown to gravitate, we expect
that during some period in the past, quantum coherence of 
gravitating systems will qualitatively alter the thermodynamics
of the cosmology.  Often, the onset of the importance of quantum
effects in gravitation is taken to be at the Planck scale.  However,
as is the case with Fermi degenerate stars, this need not be
true of the cosmology as a whole.  By quantum coherence, we refer to
the entangled nature of quantum states for space-like separations. 
This is made evident by superluminal correlations (without the
exchange of signals) in the observable behavior of such quantum states. 
Note that the exhibition of quantum coherent behavior for gravitating systems does
not require the quantization of the gravitation field.

The (luminal) horizon problem for present day cosmology examines the reason for the large
scale homogeneity and isotropy of the observed universe.  The present age of the
universe can be estimated from the Hubble scale to be
$H_o t_o \cong 0.96 \Rightarrow t_o \cong 13.2 \times 10^9$ years=
$4.16 \times 10^{17}$ seconds.
If the size of the observable
universe today is taken to be of the order of the Hubble scale ${c \over H_o} \approx 10^{28}cm$,
then if the universe expanded from the Planck scale, its size at that scale would have been of the order $\sim 10^{-4}cm$
at that time.  Since the Planck length is of the order $L_P \sim 10^{-33}cm$, then there would be
expected to be $(10^{29})^3 \sim 10^{87}$ causally disconnected 
(for luminal signals) regions in the sky.  Further, examining the
ratio of the present conformal time $\eta_o$ with that during recombination ${\eta_o \over \eta_*} \sim 100$,
the subsequent expansion is expected to imply that light from the cosmic microwave background would
come from $100^3=10^6$ disconnected regions.  Yet, angular correlations of the fluctuations have been
accurately measured by several experiments\cite{WMAP}. 

Our approach is to start from well understood macrophysics and end at the onset of microcosmology.
We refer to this period as gravitational de-coherence.  The FRW scale factor is used to compare
cosmological scales with those microscopic quantum scales we are familiar with, which define the
lengths of rulers, ticks of clocks, mass of particles, and temperatures of thermodynamic systems. 
We will insist that our calculations not depend on the present particle horizon scale, which is an accident of
history.  It will be argued that the equilibration of microscopic interactions can only occur post-decoherence.
Global quantum coherence prior to this period solves the horizon problem, since quantum correlations are
in this sense supraluminal.

Present data examining the luminosities of distant Type Ia supernovae, which have an
understood time and frequency dependency,
indicate clearly that the rate of expansion of the universe has been accelerating for several billion years\cite{TypeIa}.
This conclusion is independently confirmed by analysis of the
Cosmic Microwave Background radiation\cite{PDG}.  Both results are in quantitative agreement with a
(positive) cosmological constant fit to the data.  Our interpretation of this cosmological ``dark energy"
will be due to the vacuum energy of a quantum coherent cosmology. 

One physical system in which vacuum energy density directly manifests is the Casimir effect\cite{Casimir}.
Casimir considered the change in the vacuum energy due to the placement of
two parallel plates separated by a distance $a$.  He calculated an energy per unit
area of the form
\be
{ {1 \over 2} \left ( \sum_{modes} \hbar c k_{plates} -
\sum_{modes} \hbar c k_{vacuum}  \right ) \over A} \: = \:
- {\pi^2 \over 720} {\hbar c \over a^3}
\ee
resulting in an attractive force of given by
\be
{F \over A} \: \cong \: {- 0.013 dynes \over (a / micron)^4  } cm^{-2},
\ee
independent of the charges of the sources. 
Lifshitz and his collaborators\cite{Lifshitz} demonstrated that the Casimir force
can be thought of as the superposition of the van der Waals attractions
between individual molecules that make up the attracting media.  This allows
the Casimir effect to be interpreted in terms of the zero-point motions of the
sources as an alternative to vacuum energy.  Boyer\cite{Boyer} and others subsequently
demonstrated a repulsive force for a spherical geometry of the form
\be
{1 \over 2} \left ( \sum_{modes} \hbar c k_{sphere} -
\sum_{modes} \hbar c k_{vacuum}  \right )  \: = \:
{ 0.92353 \hbar c \over a   }.
\ee
This means that the change in electromagnetic vacuum energy \textit{is}
dependent upon the geometry of the boundary conditions.  Both
predictions have been confirmed experimentally.

The introduction of energy density $\rho$ into Einstein's equation introduces a preferred rest
frame with respect to normal energy density.  However, as can be seen in the Casimir effect,
the vacuum need not exhibit velocity dependent effects which would break Lorentz invariance.  Although
a single moving mirror does experience dissipative effects from the vacuum due to its motion,
these effects can be seen to be of 4th order in time derivatives\cite{Genet}. 

Another system which manifests physically measurable effects due to zero-point energy is liquid $^4$He. 
One sees that this is the case by noting that atomic radii are related to atomic volume $V_a$
(which can be measured) by $R_a \sim V_a ^{1/3}$.  The uncertainty relation gives momenta of the order
$\Delta p \sim \hbar / V_a^{1/3}$.  Since the system is non-relativistic, we can estimate the zero-point kinetic
energy to be of the order $E_o \sim {(\Delta p)^2 \over 2 m_{He}} \sim {\hbar ^2 \over 2 m_{He} V_a ^{2/3}}$. 
The minimum in the potential energy is located around $R_a$, and because of the low mass of $^4$He, the
value of the small attractive potential is comparable to the zero-point kinetic energy.  Therefore, this bosonic
system forms a low density liquid.  The lattice spacing for solid helium would be expected to be even
smaller than the average spacing for the liquid.  This means that a large external pressure is necessary to
overcome the zero-point energy in order to form solid helium.

Applying this reasoning to relativistic gravitating mass units with quantum coherence within the
volume generated by a Compton wavelength $\lambda_m ^3$, the zero point momentum is expected
to be of order $p \sim {\hbar \over V^{1/3}} \sim {\hbar \over \lambda_m}$.  This gives a zero point
energy of order $E_0 \approx \sqrt{2} m c^2$.  If we estimate a mean field potential from the
Newtonian form $V \sim -{G_N m^2 \over \lambda_m}=-{m^2 \over M_P ^2}mc^2<<E_0$, it is evident that
the zero point energy will dominate the energy of such a system.

For the reverse time extrapolation from the present, we adopt the
currently acccepted values\cite{PDG}\cite{PDGconstants} for the cosmological parameters involving
dark energy and matter:
\begin{equation}
h_0 \cong  0.73; \ \ \Omega_{\Lambda} \cong 0.73; \ \ \Omega_M \cong 0.27 .
\end{equation}
Here $h_0$ is the normalized Hubble parameter. Note that this
value implies that the universe currently has the critical energy
density $\rho_c=5.6 \times 10^{-4} Gev \ cm^{-3}$. We
can make the backward time extrapolation with confidence using
known physics in the customary way back to the electro-weak
unification scale $\sim 100 \ Gev$, with somewhat less confidence
into the quark-gluon plasma then encountered and beyond the top
quark regime, and expect that --- unless unexpected
new particles and/or new physics are encountered
---we can continue up to an order of magnitude higher energies
with at most modest additions to the particle spectrum. In this
radiation-dominated universe this backward extrapolation (which
taken literally {\it must} terminate when the
Friedman-Robertson-Walker scale factor $R(t)$ goes to zero and its
time rate of change $\dot R(t)$ goes to infinity) is guaranteed to
reach the velocity of light  $\dot R(t_c)=c$  at some finite time
$t_c$ when the scale factor $R(t_c))$ still has a small, but
finite, value.

As we discuss more carefully below, using our extrapolation beyond
the limit just established  (i.e.  $\dot R(t_c)=c$) would seem to
conflict with our basic methodological assumption that we invoke
no unknown physics. It is true that {\it as a metric theory of
space-time} the curved space-times of general relativity used in
the homogeneous and isotropic cosmological models we employ are
not restricted in this way. However, if we wish drive these models
by mass-energy tensors derived from either particulate or
thermodynamic models relying on some equation of state, and hence
the hydrodynamics of some form of matter, we must not use them in
such a way as to allow causal signaling at speeds greater than
$c$ by non-gravitational interactions. The exception to this
stricture which {\it is} allowed by known physics is that {\it
coherent quantum systems} have  supraluminal {\it correlations}
which cannot be used for supraluminal {\it signaling}.
Consequently, we are allowed --- as we assert in this paper --- to
start our examination of the universe at the $\dot R=c$ boundary
if it is a fully coherent quantum system. Note that the beautiful
experiments by Overhauser and collaborators already
cited\cite{Over74,Over75} justify our invocation of such systems
when they are primarily (or even exclusively) dependent on
gravitational interactions. Thus we claim that it is consistent to
start our cosmology with the cosmological {\it decoherence} of a
quantum system at the $\dot R=c$ boundary. This quantum
decoherence process is discussed in detail in Section 3:{\it
Dark Energy De-coherence}.

In Section 2:{\it Motivation} we examine an earlier
paper\cite{Noyes03} which gave cosmological reasons why $\sim 5
Tev$ might be the threshold for new physics. This paper was based
on much earlier work by E.D.Jones\cite{Jones90s,Jones97,Jones02}
and a more recent collaboration with L.H.Kaufmann and
W.R.Lamb\cite{NKLL03}. We find that, in contrast to Jones' {\it
Microcosmology} which starts with an expansion from the Planck
scale, we can identify the transition from speculative physics to
a regime which can be reached with some confidence by the backward
extrapolation from the present already assumed above. This section
shows that the identification of the critical transition with the
de-coherence of a cosmological gravitationally quantum coherent
system allows us to recover the semi-quantitative results of the
Jones theory {\it without} having to introduce speculative
physics.  However, in the development of these results, we
were motivated to re-examine the problem from a fresh perspective,
as is done in the following section.

In Section 3:{\it Dark Energy De-coherence} we will motivate
our explanation of the dark energy driving the observed acceleration
of the cosmology as the gravitational vacuum energy density
of a zero temperature Bose condensate.  Here we are able to
more quantitatively reproduce the results of the prior section
in an independent manner.  Prior to a sub-luminal
rate of expansion of the FRW scale factor, we assert that
only gravitational and quantum coherence properties are
relevant to the dynamics of the expanding cosmology.
We will develop a single parameter model in terms of the
cosmological constant, and use this to predict a mass 
and temperature scale for decoherence.  We give
an argument to support our assumption that the
condensate remains at zero temperature during the
pre-coherent phase of the cosmology.
We will end by examining the expected amplitudes of
density fluctuations if such fluctuations are the result
of dark energy de-coherence.

Finally, in Section 4:{\it Discussion and Conclusions}, we will
discuss the nature of cosmological dark energy, especially
with regards to the distant future.  It is especially interesting
to question the constancy of vacuum energy density after
a future gravitational re-coherence event $\dot{R} \geq c$. 
Some thoughts on our present and future efforts will be given.

\section{Motivation}

\subsection{Jones' Microcosmology \label{JonesMicro}}
\indent

Our present work originated in the re-examination of a paper by one of
us\cite{Noyes03} emphasizing the likelihood of some threshold for
new physics at $\sim 5 \ Tev$. This in turn was based on a
discussion with E.D. Jones in 2002\cite{Jones02}; our
understanding of this discussion and his earlier
ideas\cite{Jones90s,Jones97} has been published by us in
collaboration with L.H.Kaufmann and W.R.Lamb\cite{NKLL03}.
Briefly, Jones envisages an extremely rapid (``inflationary")
expansion from the Planck scale (i.e. from the Planck length $L_P
={\hbar \over M_P c} \cong 1.6 \times 10^{-33} cm$, where $M_{Pk} = [{\hbar
c\over G_N}]^{{1\over 2}} \cong 2.1 \times 10^{-8} kg \cong 1.221 \times 10^{19} GeV/c^2$, 
and $G_N={\hbar c \over M_P^2}$ is Newton's
gravitational constant) to a length scale $R_{\epsilon} \sim
{1\over \epsilon}$. For the reader's convenience, will will also display the
Planck time $T_P = {L_P \over c} \cong 5.4 \times 10^{-44} sec$ and the
Planck temperature $\theta_P={M_P c^2 \over k_B} \cong 1.4 \times 10^{32} \, ^o K$.
Unless necessary for clarity, we will generally choose
units such that $\hbar=1, c=1, k_B =1$.  This expansion, whose details are not
examined, is characterized by the dimensionless ratio 
\be
\mathcal{Z}_{\epsilon}\sim {R_{\epsilon}\over L_P}\sim{M_P\over \epsilon} .
\ee

When this expansion has occurred, the virtual energy which drives
it makes a thermodynamic equilibrium transition to normal matter
(i.e. dark, baryonic and leptonic, electromagnetic,...) at a
mass-energy scale characterized by the mass parameter $m_{\theta}$
and length scale ${1\over m_{\theta}}$. Jones uses the energy
parameter $\epsilon$ as a unit of energy which he calls one {\it
Planckton} defined as one Planck mass's worth of energy
distributed over the volume $V_{\epsilon} \sim {1\over
\epsilon^3}$ measured by the scale parameter $R_{\epsilon} \sim
{1\over \epsilon}$. The virtual energy is assumed to consist of
$N_{Pk}$ Plancktons of energy $\epsilon$, corresponding to an
energy density $\sim {N_{Pk}\epsilon \over V_{\epsilon}}$. This
virtual energy makes an (energy-density) equilibrium transition to
normal matter, so that
\begin{equation}
N_{Pk}\epsilon^4 \sim m_{\theta}^4 .
\label{Jones_equilibrium}
\end{equation}

It is assumed that one Planckton's worth of energy is ``left
behind" and hence that $\sim \epsilon^4$ can be interpreted as the
cosmological constant density $\rho_{\Lambda}$ at this and
succeeding scale factors. It then can be approximately evaluated
at the present day using $\Omega_{\Lambda} \sim 0.73$ as we show in
the Sec. \ref{CosCnst}. In this sense both Jones' theory and ours can be
thought of as {\it phenomenological} theories which depend only on
a {\it single} parameter (outside of the constants from
conventional physics and astronomy). We will review in the
concluding section some additional {\it observed or
potentially observable} facts that might be {\it predicted}.

\subsection{Dyson-Noyes-Jones Anomaly}
\indent

We note that our sketch of the Jones theory as expressed by Equation \ref{Jones_equilibrium}
introduces two new parameters ($N_{Pk}$,$m_{\theta}$) which must
be expressed in terms of $\epsilon$ if we are to justify our claim
that this is a single parameter theory. These parameters refer to
a very dense state of the universe. According to our methodology,
we must be able extrapolate back to this state using only current
knowledge and known physics. Jones assumes that this dense state
can be specified by using an extension to
gravitation\cite{Noyes75} of an argument first made by
Dyson\cite{Dyson52,SSS94}. Dyson pointed out that if one goes to
more that 137 terms in $e^2$ in the perturbative expansion of
renormalized QED, and assumes that this series also applies to a
theory in which $e^2$ is replaced by $-e^2$ (corresponding to a
theory in which like charges attract rather than repel), clusters
of like charges will be unstable against collapse to negatively
infinite energies. Schweber\cite{SSS94} notes that this argument
convinced Dyson that renormalized QED can never be a {\it
fundamental theory}. Noyes\cite{Noyes75} noted that {\it any}
particulate gravitational system consisting of masses $m$ must be
subject to a similar instability and could be expected to collapse
to a black hole.

We identify $Z_{e^2}= 1/\alpha_{e^2} \cong  137$ as the number of
electromagnetic interactions which occur within the Compton
wave-length of an electron-positron pair ($r_{2m_e} =
\hbar/2m_ec$) when the Dyson bound is reached. 
If we apply the same reasoning to gravitating particles of mass
$m$ (and if we are able to use a classical gravitational form),
the parameter $\alpha_{e^2}$ is replaced by
$\alpha_m=G_Nm^2={m^2\over M_P^2}$; the parameter fixing the
Dyson-Noyes (DN) bound becomes the number of {\it gravitational}
interactions within $\hbar/mc$ which will produce another particle
of mass $m$ and is given by
\begin{equation}
\mathcal{Z}_m \sim {M_P ^2\over m^2}
\end{equation}
If this dense state with Compton wavelength $\lambda_m \sim 1/m$,
contains $\mathcal{Z}_m$ interactions within $\lambda_m$, then the
Dyson-Noyes-Jones (DNJ) bound is due to the expected transition
transition $\mathcal{Z}_mm \rightarrow (\mathcal{Z}_m+1)m$, indicating instability
against gravitational collapse due to relativistic particle
creation.

It is particularly interesting to examine the production channel
for the masses m due to $\mathcal{Z}_m$ interactions within 
$\lambda_m$. This is diagrammatically represented in Figure \ref{collapse}.
\onefigure{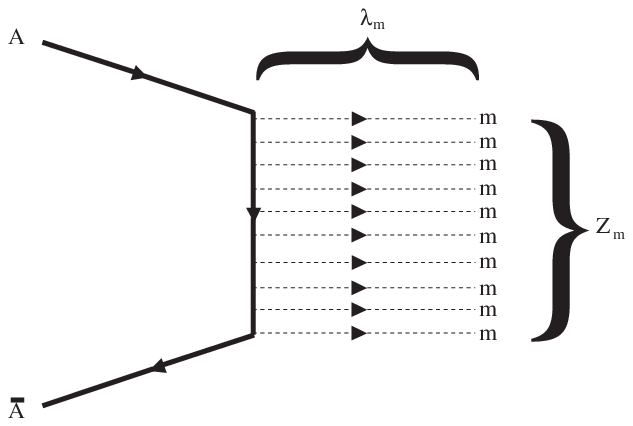}{Noyes-Jones collapse of gravitating quanta}
If there are $\mathcal{Z}_m$ scalar gravitating particles of mass m within
the Compton wavelength of that mass, a particle falling into that
system from an appreciable distance will gain energy equal to
$mc^2$, which could produce yet another gravitating mass $m$.
Clearly, the interaction becomes anomalous.

Generally, when the perturbative form of a weak interaction
becomes divergent, it is a sign of a phase transition, or a
non-perturbative state of the system (eg bound states).  One
expects large quantum correlations between systems of mass
\emph{m} interacting on scales smaller than or comparable to the
Compton wavelength of those masses.  We will assume that if the
(intensive) number of gravitational interactions (with no more
than a Planck mass worth of interaction energy per Planckton, as
defined in Section \ref{JonesMicro}) of mass units $m$ which can occur within a
region of quantum coherence is greater than the DNJ limit,
a phase transition into systems with quantum coherence scales of
the Compton wavelength of those mass units will occur. We then
expect de-coherence of subsystems of vastly differing quantum
coherence scales when the Jones transition occurs.

We could be concerned that such a concentration of mass might form
a black hole.  Although we will be primarily assuming an FRW
global geometry, if the Schwarzschild radius of a concentration of
mass is considerably less than the proper radial size of the
cosmology, one can sensibly discuss smaller regions which
approximate a Schwarzschild geometry.  For a system of a large
number $\mathcal{Z}_m$ of gravitating particles (given by the onset of the
DNJ anomaly), the Schwarzschild radius of these
masses is given by 
\be R_S \: = \: {2 G_N (\mathcal{Z}_m m) \over c^2} \: =
\: 2 \mathcal{Z}_m {m^2 \over M_P ^2}{\hbar \over mc} \: = \: 2 \lambda_m
\ee 
Therefore, such a gravitating system of masses would be
expected to be unstable under gravitational collapse. We
determine the maximum number $\mathcal{Z}_m$ of coherent interactions energy
units $m$ beyond which the system will become unstable under
gravitational collapse as 
\be R_S \: = \: \lambda_m \Rightarrow \mathcal{Z}_m
\: = {M_P ^2 \over 2 m^2} \ee 
This argument does not assume Newtonian gravitation.

A general comparison of the dependence of the Compton wavelength and
Schwarzschild radius on the mass of the system as shown in Figure \ref{LammRs}
gives some insight into regions of quantum coherence. 
\onefigure{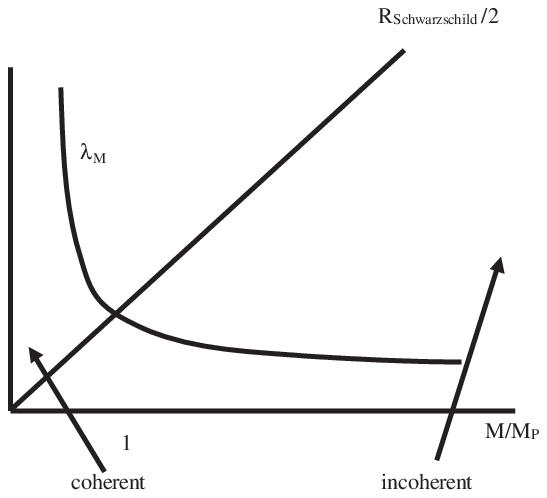}{Functional
dependence of Compton wavelength and Schwarzschild radius on
system mass} 
To the left of the point of intersection of the two
curves, the Compton wavelength is larger than the Schwarzschild
radius, and for such localized mass distributions the quantum
coherence properties are important in any gravitational
considerations.  On the other hand, for masses much larger than
the Planck mass, the gravitational distance scales are well
outside of the quantum coherence scales for isolated masses.  For
the present discussion, it is the transition region that is of
interest. An elementary particle is not expected to have a mass greater than a
Planck mass.  If the mass were greater than $M_P$, then the Compton
wavelength would be less than the Schwarzschild radius of the
particle, thereby dis-allowing coherence (for local experiments) of the particle
due to Hawking radiation, as will be discussed.

In our previous discussions\cite{NKLL03}, we considered $N_m$
particles of mass \emph{m} within $\sim \lambda_m$ and hence
$\mathcal{Z}_m={N_m(N_m -1) \over 2}$ interacting pairs. The Schwarzschild
radius in this case was considerably smaller than $\lambda_m$ 
\be
R_S \: \cong \: 2 G_N \sqrt{2 \mathcal{Z}_m}m \: \sim \: {\lambda_m \over
\sqrt{\mathcal{Z}_m}} << \lambda_m \ee 
However, in the present discussion,
$\mathcal{Z}_m$ counts the number of interactions carried by a quantum of
mass $m$.  Such interactions are expected to have a coherence length
of the order found in the propagator of a Yukawa-like particle,
$\lambda_m ={\hbar \over mc} $ The DN argument applied to
gravitation does allow us to partition 1 Planck mass worth of
energy into $\mathcal{Z}_m$ interactions within the Compton wavelength
$\lambda_m$. One way of examining Dyson's argument is to note that
if one has $\mathcal{Z}_{e^2} \cong 137$ photons of approriate energy
incident on an electron, all within its Compton wavelength
$\lambda_{m_e}$, we expect a high likelihood of pair creation. By
analogy, if there are $\mathcal{Z}_m$ coherent masses \emph{m} within
$\lambda_m$, there is high likelihood of the production of a
scalar mass $m$.

\subsection{Coherent Gravitating Matter}
\indent

As noted in the Introduction, our approach is to examine the
physical principles that we feel most comfortable using, and then
extrapolate those principles back to the earliest period in the
evolution of the universe for which this comfort level persists.
Those conclusions that can be deduced from these principles will
in this sense be model independent.  We will refrain from engaging
in constructing micro-cosmological models during earlier stages.

Following Jones we  have associated a Planckton (cf. Sec.\ref{JonesMicro}) with a
region that has quantum coherent energy of one Planck mass $M_P$.
There are expected to be many Planck units of energy within a
scale radius of the universe.  Planckta will be considered to be
internally coherent units that become incoherent with each other
during the period of de-coherence.

In general, we define $\epsilon$ as a gravitational energy scale
associated with the scale factor $R$ of the FRW metric.  On a per
Planckton basis, the average number $\mathcal{Z}_\epsilon$ of (virtual
quantum) energy units $\epsilon<M_P$ that are localizable within a
region of quantum coherence $R_\epsilon \sim 1/\epsilon$ is given
by 
\be \mathcal{Z}_\epsilon \epsilon \: = \: M_P \Rightarrow \epsilon \sim
{R_\epsilon \over L_P}= R_\epsilon  M_P\ee 
Note that for us this
understanding of the meaning of $\mathcal{Z}_{\epsilon}$ {\it replaces} the
Jones ``inflationary" definition motivated by microcosmology. This
allows us to {\it start} our  ``cosmological clock" at a {\it
finite} time calculated by backward extrapolation to the
transition. In this paper we need not consider ``earlier" times or
specify a specific $t=0$ achievable by backward extrapolation.

The localization of interactions has to be of the order $\hbar/mc$
in order to be able to use the DNJ  argument. This
allows us to obtain the number
of gravitational interactions that can occur within a Compton
wavelength of the mass $m$ which provide sufficient energy to create
a new mass or cause gravitational collapse, namely $\mathcal{Z}_m \sim {M_P
^2 \over  m^2}$. The process of de-coherence occurs when there are
a sufficient number of available degrees of freedom such that
gravitational interactions of quantum coherent states of
Friedman-Lemaitre (FL) matter-energy could have a
DNJ anomaly. We assume that a Planck mass $M_P$
represents the largest meaningful scale for energy transfer at
this boundary. A single Planckton of coherent energy in a scale of
$R_\epsilon$ will have $\mathcal{Z}_\epsilon$ partitions of an available
Planck mass of energy that can constitute interactions of this
type. \onefigure{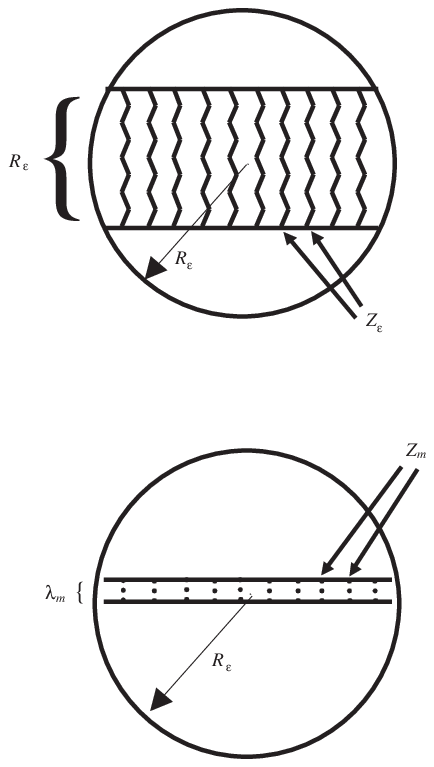}{Counting of gravitating quanta during
de-coherence} Since there is global gravitational de-coherence at
later times, the quantity $\mathcal{Z}_\epsilon$ can only be calculated
prior to and during de-coherence.

When the number of partitions of a given Planckton energy unit
equals the DNJ limit, in principle there could occur
a transition of the DNJ type involving
interaction energies equal to a Planck mass (or less than a Planck
mass at later times), thus allowing us to conclude that
de-coherence gives a mass scale from the relationship \be
\mathcal{Z}_\epsilon=\mathcal{Z}_m  \equiv \mathcal{Z} . \ee Expressed in terms of the energy
scales, this gives the fundamental equation connecting (via the
current value of $\Omega_{\Lambda}$) the observable parameter
$\epsilon$ to the mass scale at decoherence $m$ 
\be 
m^2 \approx \epsilon M_P .
\label{Jones_eqn}
\ee 
This connection is a succinct summary of Jones'
theory; henceforth we will refer to it as the \textit{Jones
equation}. Note that, viewed in this way, we no longer need the
(Jones) thermodynamic Eqn. \ref{Jones_equilibrium} to derive Eqn. \ref{Jones_eqn}. Therefore the
temperature scale we called $m_{\theta}$ need no longer be {\it
directly} identified with the particulate mass $m$ which is
associated with our {\it quantum decoherence} transition.

\subsection{Correspondence with the Measured Cosmological Constant
\label{CosCnst}}
\indent

We use the present day measurement of the cosmological density
$\rho_{\Lambda}$ to determine the energy scale $m$ of cosmological
de-coherence. The present cosmological constant energy density
$\rho_{\Lambda}$ is usually given in terms of the critical density $\rho_c$
and the reduced Hubble parameter $h$
\be \rho_c \: \equiv \: {3 H_o ^2
\over 8 \pi G_N} \: \approx \: 1.0537 \times 10^{-5} \, h^2 \,
GeV/cm^3 .\ee 
As already noted, we will take the value of the reduced Hubble
parameter to be given by $h=0.73$; current estimates of the
reduced cosmological constant energy density $\Omega_\Lambda
\equiv {\rho_\Lambda \over \rho_c} \approx 0.73$, which means that
the vacuum energy sets the de-coherence scale as
\be \rho_\Lambda
\: \approx \: 4.10 \times 10^{-6} {GeV \over cm^3} \approx 
3.13 \times 10^{-47} GeV^4  \sim \epsilon ^4
\ee 
Using the value $\hbar c \cong 1.97 \times 10^{-14} GeV \, cm$
we can immediately calculate the de-coherence scale energy and FRW
scale radius \be \epsilon \: \sim \: 10^{-12} GeV \quad \quad ,
\quad \quad R_\epsilon \: \sim \: 10^{-2} cm . \ee The Planck
energy scale and DNJ limit at this scale is given by
\be \mathcal{Z}_\epsilon \: \equiv \: {M_P \over \epsilon} \: \sim \:
10^{30} \: \sim \: \mathcal{Z}_m \ee 
The equality of the interaction factors
$\mathcal{Z}$ gives the Jones equation $m^2=\epsilon M_P$ from which
we calculate a value for the mass scale for quantum de-coherence
\be m \: \sim \: 5 \: TeV/c^2 \ee 
The number of Planck energy
units per scale volume during de-coherence is given by \be N_{Pk}
\: \sim Z^2 \: \sim \: 10^{60}. \ee 

At this point we will examine the Hubble rate equation during this transition.  If
we substitute the expected energy density into the Freedman-Lemaitre equation,
we obtain a rate of expansion given by
\be
\dot{R}_\epsilon \: \sim \: R_\epsilon \sqrt{{8 \pi G_N \over 3}( \rho_m + \rho_\Lambda) }
\: \sim \: {1 \over \epsilon} \sqrt{{1 \over M_P ^2} \mathcal{Z}_\epsilon ^2 \epsilon^4 } \: \sim \: 1.
\ee
This is a very interesting result, which implies that the transition occurs near
the time that the expansion rate is the same as the speed of light.  In 
Section 3 on dark energy de-coherence, we will develop this argument as the
primary characteristic of this transition.

\subsection{Gravitating massive scalar particle}
\indent

If the mass \emph{m} represents a universally gravitating scalar
particle, we expect the coherence length of interactions involving
the particle to be of the order of its Compton wavelength with
regards to Yukawa-like couplings with other particulate degrees of
freedom present.  If the density is greater than \be \rho_m \:
\sim \: {m \over \lambda_m ^3} \: \sim \: m^4 \ee we expect that
regions of quantum coherence of interaction energies of the order
of $m$ and scale $\lambda_m$ will overlap sufficiently over the
scale $R_\epsilon$ such that we will have a macroscopic quantum
system on a universal scale. 
\onefigure{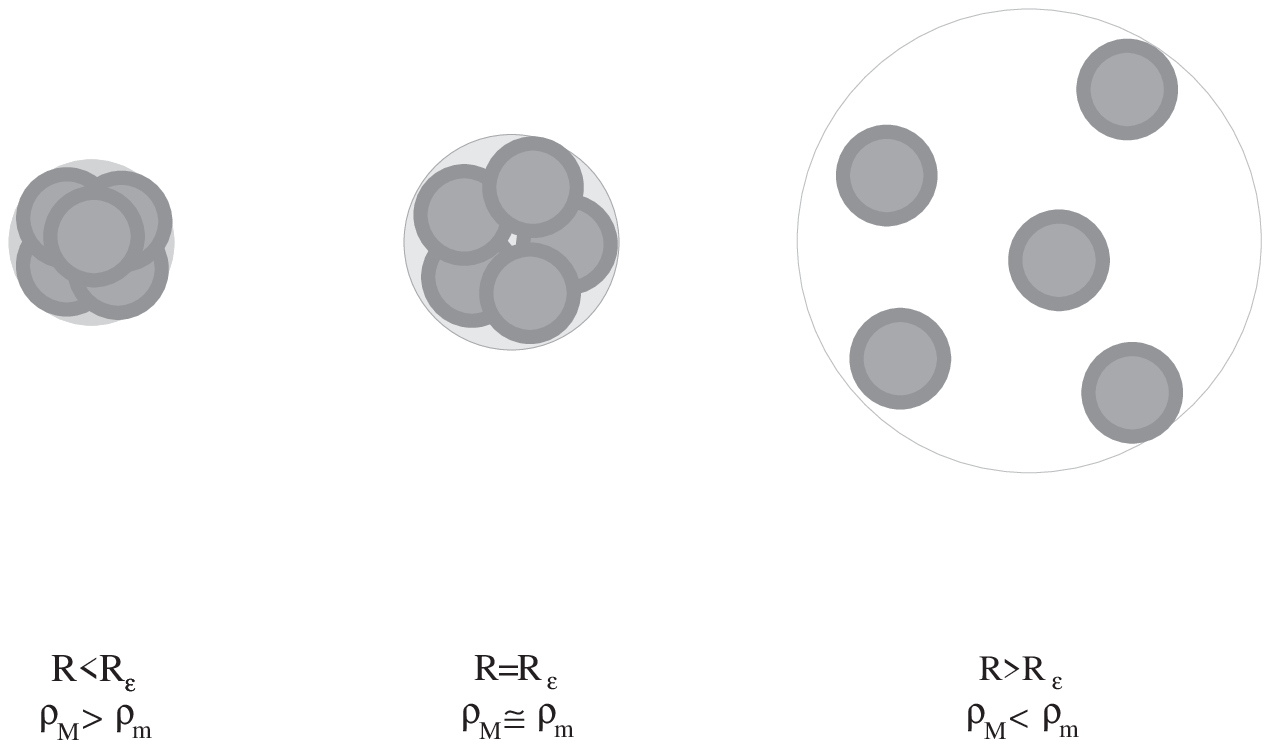}{Overlapping
regions of coherence during expansion} 
As long as the region of
gravitational coherence is of FRW scale $R<R_\epsilon$,
cosmological (dark) vacuum energy is determined by this scale.
However, when the density of FL energy becomes less than $\rho_m$,
we expect that since the coherence length of the mass m given by
its Compton wavelength is insufficient to cover the cosmological
scale, the FL energy density will break into domains of cluster
decomposed (AKLN de-coherent\cite{AKLN}) regions of local
quantum coherence. This phase transition will decouple quantum
coherence of gravitational interactions on the cosmological scale
$R_\epsilon$. At this stage (de-coherence), the cosmological
(dark) vacuum energy density $\rho_\Lambda$ is frozen at the scale
determined by $R_\epsilon$. The cosmological dark energy
contribution to the expansion rate is so small, and its coupling
to de-coherent FL energy so insignificant, that its value is
frozen at the value just prior to de-coherence given by \be
\rho_\Lambda \: \sim \: {\epsilon \over R_\epsilon ^3} \: = \:
\epsilon ^4 \ee

We should note that using the DNJ argument, if the mass \emph{m}
were engaged in active cosmological energy exchanges involving
non-gravitational microscopic interactions prior to decoherence,
then, since those interactions have considerably larger coupling
constants, it can be concluded that 
$\mathcal{Z}_g = {1 \over g^2} < \mathcal{Z}_m$.
This would mean that this interaction would have broken coherence
prior to our expected gravitational de-coherence event. We expect the mass m to
be dark during this period (as must be all other particles).

During de-coherence, we assume that the FL energy contained in
$R_\epsilon$ is given by $N_{Pk}$ Planck mass units appropriately
red-shifted to the de-coherence epoch.  This gives an intensive FL
energy density (for a spatially flat universe) of the form \be
\rho_{FL} \: \sim \: N_{Pk} \epsilon ^4 \ee Since de-coherence is
expected to occur when this density scale is given by the quantum
coherence density scale for the mass by $\rho_m$, we obtain the
following relationship between the de-coherence energy scale
$\epsilon$ and the scalar mass $m$: \be N_{Pk} \epsilon ^4 \:
\cong \: m^4 \ee This allows us to consistently relate the number
of Plancktonic energy units in the region of coherence
$R_\epsilon$ to the DN counting parameters: 
\be N_{Pk}
\cong {m^4 \over \epsilon^4}= {m^4 \over M_P ^4} 
{M_P ^4 \over \epsilon^4} = 
{\mathcal{Z}_\epsilon ^4 \over \mathcal{Z}_m ^2}=
\mathcal{Z}^2 , 
\ee 
which insures that all quantities relevant to our theory can be reduced to a single parameter 
in the Jones equation $m^2 \cong \epsilon M_P$.

It has already been suggested\cite{Noyes03} that if we identify
the Jones mass parameter $m$ with a massive, scalar gravitating
particle, this could be a candidate for particulate dark matter.
Unfortunately if we  assume that the mass m interacts {\it only}
gravitationally, such a particle would be difficult to discover in
accelerator experiments due to the extremely small coupling of
gravitational scale forces.   

We hope to be able to estimate the expected dark matter to photon number ratio
from available phenomenological data if the mass is known. 
The FL equations satisfy energy conservation $T^{\mu \nu} _{; \nu}=0$,
which implies 
$\dot{\rho}=-3H(\rho + P)$.  The first law of thermodynamics relates
the pressure to the entropy density $\dot{P}={S \over V} \dot{T}$,
which then implies an adiabaticity condition on the expansion given
by
\be
{d \over dt} \left ( {S \over V} R^3 \right ) \: = \: 0.
\ee
Assuming adiabatic expansion, we expect $g(T) \, (R T)^3$ to
be constant far from particle thesholds.  Here, $g(T)$ counts
the number of low mass particles contributing to the cosmological
entropy density at temperature $T$.  This gives a red shift in terms
of the photon temperature during a given epoch
\be
{R_o \over R} \: \equiv \: 1+z \: = \: 
(1 + z_{dust}) \left ( {g(T) \over g_{dust}} \right )^{1/3} 
{T \over T_{dust}},
\label{redshift}
\ee 
where $z_{dust}$ is defined as the redshift at equality of radiation
and pressure-less matter energy densities.
In terms of photon temperature,
we can count the average number of photons using
standard results from black body radiation
\be
{N_\gamma \over N_{\gamma \, o}} \: = \: {(R \, T)^3 \over (R_o \, T_o)^3} 
\: \cong \: {g_o \over g(T)} .
\ee

This allows us to write a formula for the dark matter - photon ratio at the
temperature of dark matter number conservation ($T_{freeze}$), in terms of its mass
and the measured baryon-photon ratio:
\be
{N_{dm} \over N_\gamma} \: \cong \: 
{\Omega_{dm} \over \Omega_{baryon}} 
{N_{b \, o} \over N_{\gamma \, o}} {m_N \over m}
{g(T) \over g_o} ,
\label{m-photon-ratio}
\ee
which gives
${N_{dm} \over N_\gamma} \: \cong \: 2.9 \times 10^{-9} {g(T_{freeze}) \over g_o}
{m_N \over m}$.

If there is DNJ collapse, one can estimate the lifetime of the
resulting black holes. These collapsed objects would be expected
to emit essentially thermal low mass quanta at a rate determined
by the barrier height near the horizon $\sim M G_N$ and the
wavelength of the quanta $\sim (M G_N)^{-1}$ giving a luminosity of order
$dM/dt \sim -1/M^2 G_N ^2$.  This can be integrated to give a
lifetime of the order 
\be t_{evaporation} \: \sim \: M^3 G_N ^2 .
\ee 
This means that a collapsed DNJ object has an approximate
lifetime of 
\be \tau_{BH} \: \sim \: {(\mathcal{Z}_m m)^3 \over M_P ^4} \:
\sim \: \mathcal{Z}_m {\hbar \over m c^2} \ee 
Substituting the expected
mass, the lifetime is expected to be $\tau_{BH} \sim 400 sec$,
which is long compared to the inverse Hubble rate $H_\epsilon
^{-1} \sim 10^{-13} sec$ during decoherence.

We can also estimate the number of low mass quanta that would
result from the evaporation.  We will examine the quantum mechanics
of massive scalar particles $g_{\mu \nu}p^\mu p^\nu =-m^2$
in a Schwarzschild metric.  
\be
ds^2 = -(1-{2 G_N M \over r}) dt^2 + {dr^2 \over 1-{2 G_N M \over r} } +
r^2 d \theta^2 + r^2 sin^2 \theta d\phi^2 .
\ee
Using tortoise coordinates $r^*$, where $r^* \equiv r+2 G_N M log \left ({r \over 2 G_N M} -1  \right )$,
the action is taken to be
\be
W= {1 \over 2} \int dt dr^* d\theta d\phi \: r^2 \left( 1 - {2 G_N M \over r}  \right ) sin \theta
\left [ {-\chi_t ^2 + \chi_{r^*} ^2 \over 1 - {2 G_N M \over r} } + m^2 \chi^2 +
{\chi_\theta ^2 \over r^2}  + {\chi_\phi ^2 \over r^2 sin^2 \theta} \right ] .
\ee
The equation of motion generated for the reduced radial function $\psi \equiv r \chi$
for stationary states is given by
\be
\psi_{r^* r^*} - \left( 1 - {2 G_N M \over r}  \right ) \left ( m^2 + {\ell (\ell +1) \over r^2} +
{2 G_N M \over r^3} \right ) \psi
= \psi_{tt} = -(m^2 + k_\infty ^2) \psi.
\ee
The effective potential barrier height is seen to be of the order of the inverse
Schwarzschild radius.  The asymptotic solution satisfies
$\psi_{r r} (r \rightarrow \infty) \cong -k_\infty ^2 \psi$, whereas the solution near the Schwarzschild
radius (for s-waves) is given by 
$\psi_{r^* r^*} (r \rightarrow 2 G_N M) \cong -(m^2 + k_\infty ^2) \psi$.  We expect that
when the temperature is above mass threshold, the particle can be radiated,
and that for temperatures above the barrier height $V_{max} \sim {0.3 \over 2 G_N M}$
the transmission rate of particles is of order ${k_\infty \over \sqrt{m^2 + k_\infty ^2 }}$.
 
Writing the luminosity and number
rate as 
\be
\begin{array}{l}
{dM \over dt} \: = \: {\eta(M) \over M^2 G_N ^2 } \\
\quad \Rightarrow \: t \: = \: {1 \over 3} {M^3 G_N ^2 \over \bar{\eta}} \\
{dN \over dt} \: = \: { \eta(M) \over M G_N  } \: \cong \:
{\bar{\eta}^{2/3} \over (3 G_N t)^{1/3}}
\end{array}
\ee 
Here $\eta(M)$ is expected to be a slowly varying
function of the temperature that counts the number
of low mass thermal states at the temperature of the black hole. 
This factor is expected to be essentially constant between
particle thresholds. 
The solution then takes the form 
\be N(M) \: \cong \: {1 \over 2}
{M^2 \over M_P ^2}. \ee
More generally, the total number of low mass quanta
resulting from evaporation from mass $M$ to mass $M'$
is expected to satisfy
\be
N(M \rightarrow M') \: \cong \: {M^2 - M'^2 \over 2 M_P^2}.
\ee
If the black hole has formed due to
DNJ collapse, substituting $M=\mathcal{Z}_m m$ gives 
\be N(\mathcal{Z}_m m) \: \cong \: 
{1 \over 2} \mathcal{Z}_m ^2 {m^2 \over M_P ^2} \cong \mathcal{Z}_m. \ee
Therefore, the intermediate quanta in the collapse are expected to
produce an essentially equal number of low mass quanta during
evaporation.

We can estimate the relative number of quanta of mass $m$
evaporated by a black hole formed by DNJ collapse.  If the mass
cannot be radiated prior to temperature $T_m$, this ratio is
given by
\be
{N_m \over N_{Total}} \: \cong \: {1 \over g(T_m)}
\left(  {M(T>T_m) \over M}  \right )^2,
\ee
where $g(T_m)$ is the number of low mass states available
for radiation at temperature $T_m$.  Substituting $M=Z_m m$,
$T_m ={1 \over 8 \pi G_N M(T_m)} \cong m $ and $g(T_m)={427 \over 4}$
gives an estimate of ${N_m \over N_{Total}}\approx 10^{-5}$ from each black
hole thermalization.  This is all that can be concluded at present
relevant to the dark matter-photon ratio during thermalization.

To summarize, our re-examination of the Jones theory has led us to
the conclude that the $\mathcal{Z}_{\epsilon} = \mathcal{Z}_m$ relation is best
interpreted in that context as the equality of the (intensive)
number of gravitational quanta of mass $m$ exchanged between all
gravitating systems between the cosmological scale $R_{\epsilon}$
and the particulate scale $\lambda_m$, when the DNJ
bound $\mathcal{Z}m\rightarrow (\mathcal{Z}+1)m$ is reached. One way of examining
Dyson's argument is to note that if one has $\mathcal{Z}_{e^2}\cong 137$
photons of approriate energy incident on an electron, all within
its Compton wavelength $\lambda_{m_e}$, we expect a high
likelihood of pair creation. By analogy, if there are $\mathcal{Z}_m$
coherent masses m within $\lambda_m$, there is high likelihood of
the production of a scalar mass m. This interpretation {\it
requires} us to be talking about {\it quantum coherent systems}
when the Jones transition from microcosmology to a universe where
we can use conventional physics and cosmology takes place. This
line of reasoning suggested to us that this transition itself {\it
must} in some sense correspond to {\it quantum decoherence} and to
the title of this paper. The consequences of pursuing this line of
thought constitute the rest of this paper.

\section{Dark Energy De-coherence}
\indent

We will now make quantitative arguments to develop the general ideas
motivated by the previous sections.  Although the arguments are
independent of those in the previous section, we will derive very
similar results.  In most of what follows we will
assume flat spatial curvature $k=0$.  Prior to the scale condition
$\dot{R}_\epsilon =c$, which we will henceforth refer to as the time
of dark energy de-coherence, gravitational influences are
propagating (at least) at the rate of the gravitational scale expansion, and
microscopic interactions (which can propagate no faster than c)
are incapable of contributing to cosmological scale equilibration.  Since
the definition of a temperature requires an equilibration of interacting
"microstates", there must be some mechanism for the redistribution
of those microstates on time scales more rapid than the cosmological
expansion rate, which can only be gravitational. 

\subsection{Dark Energy}
\indent

As we have discussed in the motivation section, we expect that 
dark energy de-coherence occurs when the FRW scale is 
$R_\epsilon \sim 1/ \epsilon$.  The gravitational dark energy scale associated with de-coherence is given by $\epsilon$, independent of the actual number of energy units $N_\epsilon$ in the scale region.  
Since the dark energy density must be represented by an intensive parameter 
which should be the same for the universe as a whole, we will express this
density as the coherent vacuum state energy density of this macroscopic quantum system.  
In the usual vacuum state, the equal time correlation function 
\begin{center}
$<vacuum|\Psi(x,y,z,t) \, \Psi(x',y',z',t)|vacuum>$
\end{center}
does not vanish for space-like separations. (For example, for massless scalar fields, 
this correlation function falls off with the inverse square of the distance between the points).  
Since we assume no physical distinction between spatially separated points, 
our correlation functions would be expected to be continuous, 
with periodic boundary conditions defined by the cosmological scale factor.  
Given a cosmological scale factor $R_\epsilon$, periodic boundary conditions on long range 
(massless or low mass) quanta define momentum quantization in terms of this maximum wavelength.  
The energy levels associated with these quanta would satisfy the usual condition
\be
E_{N_\epsilon} \: = \: \left( N_\epsilon + {1 \over 2} \right ) \, \hbar \omega
\: = \: \left( 2 N_\epsilon + 1 \right ) \epsilon .
\ee

This is associated with quanta of wavelength of the order of the cosmological scale factor
with vacuum energy density, given by
\be
\rho_\Lambda \: \equiv \: {\epsilon \over (2 R_\epsilon) ^3} \: = \:
\left ( {k_\epsilon \over 2 \pi} \right )^3 {\sqrt{m_{condensate} ^2 + k_\epsilon ^2} \over 2}
\label{rholambda}
\ee
for a translationally invariant universe with periodicity scale $2 R_\epsilon$.
In effect, this provides the infrared cutoff for cosmological quantum coherent
processes,
\be
k_\epsilon \: = \: {2 \pi \over \lambda_\epsilon } \: = \: {\pi \over R_\epsilon}.
\ee
We will begin by examining a massless condensate.

The vacuum energy scale associated with a condensate of gravitationally
coherent massless quanta is given by
\be
\epsilon \: = \: {1 \over 2} \hbar \omega_\epsilon \: = \: {1 \over 2} \hbar k_\epsilon c
\: = \: { \pi \over 2} {\hbar c \over R_\epsilon},
\ee
and the vacuum energy density for such massless quanta is
\be
\rho_\Lambda \: = \: {\epsilon ^ 4 \over ( \pi) ^ 3}.
\ee

We might inquire into the nature of the dark energy, in the sense as to whether it is geometric or quantum mechanical in origin.  From the form of Einstein's equation
\be
\mathcal{R}_{\mu \nu} - {1 \over 2} g_{\mu \nu} \mathcal{R} \: = \:
{8 \pi G_N \over c^4} T_{\mu \nu} + \Lambda g_{\mu \nu} \quad \quad ,
\quad \quad \left( \Lambda g^{\mu \nu} \right )_{; \nu}=0
\ee
if the term involving the cosmological constant should most naturally appear on the left hand side of the equation, we would consider it to be geometric in origin.  If the cosmological term is geometric in origin, we would expect it to be a fundamental constant of the cosmology which scales with the FRW/FL cosmology consistently with the vanishing divergence of the Einstein tensor.  However, if the previous arguments are interpreted literally, the dark energy density freezes out to a constant determined by the period of last quantum coherence with the FL energy density and the onset of the equilibration of states involving microscopic non-gravitational interactions, supporting its interpretation as a gravitational quantum vacuum energy density.  This means that it is fixed by a physical condition being met, and thus would not be a purely geometric constant.

\subsection{Rate of Expansion during De-Coherence--massless condensate}
\indent

We will next examine the rate of the expansion during the period of de-coherence. 
We will make use of the Friedmann-LeMaitre
(FL)/Hubble equations, which relates the expansion rate and acceleration to the densities
\be
\left ( {\dot{R} \over R} \right ) ^2 \: = \: {8 \pi G_N \over 3} 
\left ( \rho + \rho_\Lambda  \right ) \, - \, {k \over R^2} ,
\label{Hubble_eqn}
\ee
\be
{\ddot{R} \over R} \: = \: 
-{4 \pi G_N \over 3} (\rho + 3 P - 2 \rho_\Lambda) ,
\label{acceleration}
\ee
where we have written
\be
\rho_\Lambda \: = \: {\Lambda \over 8 \pi G_N},
\ee
and $\rho$ represents the FL energy density.
The only scale dependent term in this equation involves the spatial curvature $k$. 
If $k$ is non-vanishing, we have no reason to assume that any scale other than 
$R_\epsilon$ at de-coherence determines the cosmological scale.  
In our discussion, the dark energy density will have negligible contribution to
the FL expansion during de-coherence, but will become significant as the FL energy density 
$\rho$ decreases due to the expansion of the universe.

It is unclear whether one can speak of causal horizons and causal communications in the 
usual ways prior to the period of de-coherence, since the scale expansion rate is larger than c. 
 Assuming that local inertial physics satisfies the principle of equivalence with a limiting velocity of c, 
it would be difficult to extrapolate the type of physics we do presently into a domain with local 
expansion rates greater than c.  Only the FRW gravitation interacts with rates which can equilibrate states 
defining a thermal system in this domain, since the other interactions cannot have 
super-luminal exchanges, only super-luminal quantum correlations.  
If the expansion rate is super-luminal $\dot{R}>c$, 
scattering states cannot form decomposed (de-coherent) clusters of the type described in reference \cite{AKLN}.
We see from the above discussion that, assuming the validity of an FL universe back to the 
stage of de-coherence, our usual ideas of microscopic causality become obscure beyond this period.

Since we find the expansion rate equation $\dot{R}_\epsilon = c$ a compelling argument for the quantitative 
description of gravitational de-coherence, it is this relationship that we will use to determine the form for the energy density 
during dark energy de-coherence $\rho_{FL}$, which counts the number of gravitating quanta above 
vacuum energy in the condensed state.  The Hubble equation takes the form
\be
H_\epsilon ^2 =
\left( {c \over R_\epsilon} \right )^2 \: = \: {8 \pi G_N \over 3} \left ( \rho_{FL} + \rho_\Lambda \right ) - 
{k c^2 \over R_\epsilon ^2}
\: = \: {8 \pi G_N \over 3} \left ( 2 N_\epsilon + 1 \right ) \rho_\Lambda - {k c^2 \over R_\epsilon ^2}.
\label{decoherence}
\ee
We see that $2 N_\epsilon$ counts the number of Jones-Planck energy units per scale factor in
the pre-coherent universe (referred to by Jones as $N_{Planckton}$),
and it defines the ratio of normal to dark energy density during de-coherence.
 
If this condition is to describe the onset of dark energy de-coherence, we can see that a so called
``open" universe ($k=-1$) is excluded from undergoing this transition.  In this case,
the cosmological constant term in equation \ref{Hubble_eqn} already excludes a solution
with $\dot{R}_\epsilon \leq c$.  

Likewise, for a ``closed" universe that is initially radiation dominated, 
we can compare the scale factors corresponding to $\dot{R}_\epsilon=c$ 
and $\dot{R}_{max}=0$. 
From the Hubble equation
\be
{c^2 \over R_{max}^2} \: = \: { 8 \pi G_N \over 3} (\rho + \rho_\Lambda) \: \cong \:
{ 8 \pi G_N \over 3} \rho_\epsilon {R_\epsilon ^4 \over R_{max} ^4} \Rightarrow
R_{max}^2 \: \cong  \: 2 R_\epsilon ^2 .
\ee
Clearly, this closed system never expands much beyond the transition scale.  
For this reason, henceforth we will only consider flat spaces.

We will assert that de-coherence cannot occur prior to $\dot{R}=c$ since incoherent decomposed
clusters \cite{LMNP} cannot be cosmologically formulated.  Using the equation
\be
\left( {c \over R_\epsilon} \right )^2 \: = \: {8 \pi G_N \over 3} \left ( 2 N_\epsilon + 1 \right ) \rho_\Lambda
\ee
and the form of $\rho_\Lambda$ from equation \ref{rholambda}
we can directly determine number of quanta in the condensed state
\be
N_\epsilon \: = \: {1 \over 2} \left ( {3 \over 2} {M_P ^2 \over \epsilon ^2} - 1  \right ) \: \cong \: 
{3 \over 4} \mathcal{Z}_\epsilon ^2
\ee
where, as before $\mathcal{Z}_\epsilon \equiv {M_P \over \epsilon}$.
The energy density during dark energy decoherence is therefore given by
\be
\rho_{FL} \: = \: 2 N_\epsilon \rho_\Lambda \: \cong \:
{3 \over 2 \pi ^3} M_P ^2 \epsilon ^2 .
\label{rhoFL}
\ee

\subsection{Phenomenological correspondence--massless condensate}
\indent

To make correspondence with observed cosmological values, we will
utilize parameters obtained from the Particle Data Group\cite{PDG}. 
The value for the critical density is given by
\be
\rho_c \: \equiv \: {3 H^2 \over 8 \pi G_N} \: \cong \: 5.615 \times 10^{-6} \: GeV / cm^3 \: \cong \:
4.293 \times 10^{-47} GeV^4.
\ee
The cosmological dark energy density parameter
$\Omega_\Lambda = {\rho_\Lambda \over \rho_c}$ is taken to have the
value $\Omega_\Lambda \cong 0.73$.  We will therefore use the value
\be
\rho_\Lambda \: \cong \: 4.099 \times 10^{-6} GeV / cm^3 \: \cong \:
3.134 \times 10^{-47} GeV^4 .
\ee
This gives a dark energy scale and FRW scale given by
\bea
\epsilon \: \cong \: 5.58 \times 10^{-12} \: GeV \\
R_\epsilon \: \cong \: 5.54 \times 10^{-3} \: cm.
\eea
The Friedman-Lemaitre energy density from equation \ref{rhoFL} is then given by
\be
\rho_{FL} \: \cong \: 2.93 \times 10^{55} \: GeV/cm^3 \: \cong \:
2.24 \times 10^{14} \: GeV^4,
\ee
with the Planck energy partition $\mathcal{Z}_\epsilon$ and number of ``gravons"
$N_\epsilon$ at dark energy de-coherence given by
\bea
\mathcal{Z}_\epsilon \: \cong \: 2.19 \times 10^{30} \\
N_\epsilon \: \cong \: 3.59 \times 10^{60}.
\eea
By \textit{gravons} we will mean gravitationally coherent Bose states.
The coherent mass density scale ${m \over \lambda_m ^3}=m^4$ corresponding to the
FL density $\rho_{FL}$ is given by $m \sim 3800 \: GeV/c^2$.  However, we will
obtain a precise determination of the coherent mass scale in terms of the UV
cutoff scale of the gravitational dynamics when we discuss the thermal
ground state shortly.

We will next estimate the time of dark energy de-coherence assuming
a radiation dominated expansion prior to this period. 
Although we are dubious about using a standard radiation dominated
equation of state prior to dark energy de-coherence, we can get some
feeling for the time scale of this transition.  The period after de-coherence
is radiation dominated until the dust driven epoch, with the scale factor satisfying
\be
R(t) \: = \: R_\epsilon \left ( {t \over t_\epsilon} \right ) ^{1/2} .
\ee
This means that ${\dot{R}(t) \over R(t)}={1 \over 2 t}$, resulting in an estimate
for the time
\be
t_\epsilon \: \cong  \: {R_\epsilon \over 2 c} \: \cong \: 9.30 \times 10^{-14} \, sec.
\ee
This also gives a Hubble rate of
\be
H_\epsilon \: = \: {c \over R_\epsilon} \: \cong \: 
5.38 \times 10^{12} /sec \: \cong \: (1.86 \times 10^{-13} sec)^{-1}.
\ee

This Hubble rate gives a minimal lifetime for any gravitating
mass scale $m$ that can equilibrate during de-coherence.  If the mass
is to have meaningful coherence during the period of de-coherence, its
lifetime in the thermal bath must be of an order greater than the inverse
Hubble rate.  This means that
\be
\tau_m \: > \: {1 \over H_\epsilon } \: \sim \: 10^{-13}sec .
\ee
If the mass scale is associated with the Higgs scalar of
the symmetry breaking, this mass could ONLY couple to electro-weak
bosons to generate mass, since the Yukawa coupling to masses
comparable to the top quark mass would give a width well in excess of
this scale.

The estimates are only slightly modified for vector and tensor gravons. 
Substituting spin degeneracy corresponding to the particle type, the
scale factor of de-coherence becomes 
$R_V \cong 7.3 \times 10^{-3}cm$ for vector gravons, and
$R_T \cong 8.3 \times 10^{-3}cm$ for tensor gravons, with the other
calculated quantities varying accordingly.  We will assume scalar
quanta for our further calculations.

\subsection{Thermal Ground State}
\indent

For a hot, thermal system, the ground state is not that
state which satisfies $\hat{N}|0>=0$ for all modes (zero occupation),
but instead is constructed of a thermal product of occupation number
states, weighted by a density matrix.  Unlike the zero occupation
number state, this ground state need not generally be time translationally
invariant.  Examining the low energy modes at high temperatures,
the thermally averaged occupation of those modes
$<\hat{N}_n> \cong {k_B T \over E_n}$
demonstrates large numbers of low energy massless quanta, giving these
modes a large number of degrees of freedom.  For our system,
there are natural infrared and ultraviolet cutoffs provided by the
macroscopic scale $k_\epsilon$ and microscopic scale $m$. 
We expect macroscopic gravitational physics involving gravitating
masses $m$ to be cutoff for momenta $k_{UV} \sim m$.

It is of interest to calculate the energy of the zero-occupation number
state using these cutoffs,
\be
\hat{H} |0,0, ... ,0> \: = \: \sum_{\vec{k}=\vec{k}_{\epsilon}}^{\vec{k}_{UV}} 
{1 \over 2} \hbar c \hat{k} \: |0,0, ... ,0>.
\ee
Inserting the density of states to approximate the sum gives
an energy density of the form
\be
{E_{|0,...,0>} \over V }
\: \sim \: {1 \over (2 \pi)^3} \int _{k_{\epsilon}} ^{m}
{1 \over 2} \hbar c k \: 4 \pi k^2 dk \: \sim \:  m^4 - \epsilon^4 
\: \sim \: \rho_{FL},
\ee
which is essentially the Jones equilibrium condition equation \ref{Jones_equilibrium}. 
This means that the vacuum energy density corresponding to zero
occupancy of the gravitational modes corresponds to the
energy density of the normal gravitating matter just after decoherence
if the ultraviolet cutoff of the long range modes in the superfluid is
chosen to be the  mass scale $m$.  We will therefore proceed 
recognizing that the mass scale provides an ultraviolet gravitational cutoff for
the decoherent cosmology.

As has been previously discussed, the vacuum energy associated
with the condensate is given by $\rho_\Lambda=\epsilon/(2 R_\epsilon)^3$,
which for a massless condensate gives $\rho_\Lambda=\epsilon^4/ \pi^3$. 
However, once the expansion rate is sub-luminal, global gravitational
coherence is expected to be broken due to interactions that
propagate at the speed of light.  This means that all available modes
must thereafter be included in calculations of the vacuum energy.  As is the
case with superfluids, we will assume that there is an ultraviolet
cutoff associated with the (scalar) mass scale $m$ with coherence length
$\lambda_m = \hbar / mc$.  The vacuum energy density associated
with this cosmology transition during dark energy de-coherence
from $\rho_\Lambda$ (at pre-coherence) to $\rho_{vac}$ (at de-coherence) is given by
\be
\begin{array}{r}
\rho_{vac} \: = \: {E_{vacuum} \over V_\epsilon} \: = \: 
\int {g_m \over 2} \hbar c |\vec{k}| {d^3 k \over (2 \pi)^3} \\
{\hbar c \over (2 \pi)^2} \int_{k_\epsilon} ^{k_m} k^3 dk \: = \:
{1 \over 4} {\hbar c \over (2 \pi)^2} (k_m ^4 - k_\epsilon ^4) ,
\end{array}
\ee
where $k_m = {2 \pi \over \lambda_m} = {2 \pi m c \over \hbar}$,
$k_\epsilon = {\pi \over R_\epsilon} = {2 \epsilon \over \hbar c}$, and the
spin degeneracy $g_m$ will be taken to be unity.  Therefore,
the vacuum energy at de-coherence is taken to be 
\be
\rho_{vac} \: = \: \pi^2 \left ( m^4 - \left (  {\epsilon \over \pi} \right ) ^4  \right )
\: \cong  \: \pi^2 m^4 .
\ee
If we presume minimal parametric input to this model, then this vacuum
energy thermalizes as the FL energy density in equation \ref{decoherence} for the cosmology
$\rho_{FL}=\rho_{vac}$, giving a relationship for the mass scale of 
dark energy de-coherence
\be
m^4 \: = \: {3 \over 2 \pi^5} M_P ^2 \epsilon ^2 .
\ee
This gives an expected mass scale given by
\be
m \: \cong \: 2183 GeV/c^2 .
\label{m_estimate}
\ee
If $g_m$ is the spin degeneracy associated with $m$, then
the left hand side of equation \ref{m_estimate} is modified
by a factor of $g_m ^{1/4}$.

The existence of a gravitational mass associated
with de-coherence introduces the possibility that the
vacuum energies should be calculated in terms of
the vacuum states of this mass rather than in terms of
the long range excitations (gravons) treated previously. 
More generally, the mass scale associated with the condensate
need not be the same as that of the ultraviolet cutoff, which
introduces yet another mass scale.  For instance, the cutoff mass $m$
could be associated with a dark matter mass, while the condensate
mass could be associated with the symmetry breaking scale. 
In the present context, we will
associate these two scales as identical.
Thus far, there is nothing in our discussion preventing the use of
vacuum energy as
\be
\epsilon_m \: \equiv \: {1 \over 2} \sqrt{m^2 + k_\epsilon ^2} \: = \:
{1 \over 2} \sqrt{m^2 + \left ( {\pi \over R_\epsilon} \right ) ^2}.
\ee
The post-decoherence vacuum energy then is given in general by
\be
\rho_{vac} \: = \: {E_{vacuum} \over V_\epsilon} \: = \: 
\int_{k_\epsilon} ^{k_m} g_m {\sqrt{m^2 + |\vec{k}|^2 } \over 2} {4 \pi |\vec{k}|^2 d k \over (2 \pi)^3} ,
\ee
which can be used to solve for the mass $m$
self-consistently by setting $\rho_{vac}=\rho_{FL}$.

\subsection{Phenomenological correspondence--massive condensate}
\indent

We will recalculate the phenomenological parameters for a pre-coherent condensate
of massive particles of mass $m$.  The self-consistent mass that satisfies the condition
$\rho_{vac}=\rho_{FL}$ is given by $m \cong 19.74 GeV$.
The dark energy scale and FRW scale is given by
\bea
\epsilon \: \cong \: 9.87 \: GeV \\
R_\epsilon \: \cong \: 67.0 \: cm.
\eea
The Friedman-Lemaitre energy density from equation \ref{rhoFL} is then given by
\be
\rho_{FL} \: \cong \: 2.01 \times 10^{47} \: GeV/cm^3 \: \cong \:
1.54 \times 10^{6} \: GeV^4,
\ee
with the Planck energy partition $\mathcal{Z}_\epsilon$ and number of condensate particles
$N_\epsilon$ at dark energy de-coherence given by
\bea
\mathcal{Z}_\epsilon \: \cong \: 1.24 \times 10^{18} \\
N_\epsilon \: \cong \: 2.45 \times 10^{52}.
\eea

The time estimate for a radiation dominated cosmology is given by
\be
t_\epsilon \: \cong  \: {R_\epsilon \over 2 c} \: \cong \: 1.13 \times 10^{-9} \, sec.
\ee
This gives a Hubble rate of
\be
H_\epsilon \: = \: {c \over R_\epsilon} \: \cong \: 
4.45 \times 10^{8} /sec \: \cong \: (2.25 \times 10^{-9} sec)^{-1}.
\ee

Again, the estimates are only slightly modified for vector and tensor masses. 
Substituting spin degeneracy corresponding to the particle type, the
scale factor corresponding to de-coherence becomes 
$R_V \cong 62.0 cm$ for vector masses $m \cong 15.6$ GeV, and
$R_T \cong 59.8 cm$ for tensor masses $m \cong 14.0$ GeV, with the other
calculated quantities varying accordingly.  We will assume scalar
masses for our calculations.

\subsection{Thermalization}
\indent

We will next examine the thermalization of the coherent gravitating cosmology
into the familiar particulate states.  De-coherence is presumed to occur
adiabatically into a radiation dominated cosmology.  For each low
mass particle state, the standard black body relationships
are satisfied:
\be
{U \over V}\: = \: g {\pi ^2 \over 30} \left( {k_B T \over \hbar c} \right ) ^3
\, k_B T \: = \: \rho \, c^2 \: = \: 3 P
\ee
\be
{S \over V}\: = \: g {2 \pi ^2 \over 45} k_B \left( {k_B T \over \hbar c} \right ) ^3
\ee
\be
{N \over V}\: = \: g^* {\zeta (3) \over \pi^2} \left( {k_B T \over \hbar c} \right ) ^3
\ee
where the statistical factors are given by
\be
g/ \# \,spin \, states = \left \{ 
\begin{array}{l} 1 \quad \textnormal{bosons}\\
{7 \over 8} \quad \textnormal{fermions} \end{array} \right .
\quad , \quad 
g^*/ \# \, spin \, states = \left \{ 
\begin{array}{l} 1 \quad \textnormal{bosons}\\
{3 \over 4} \quad \textnormal{fermions} \end{array} \right . 
\ee

The temperature of radiation with density $\rho_{FL}$ is given by
\be
\rho_{FL} \: = \: g(T_\epsilon) {\pi^2 \over 30} 
{(k_B T_\epsilon)^4 \over (\hbar c)^3 } \: \Rightarrow \:
T_\epsilon \: \cong \: {T_{crit}  \over g^{1/4}(T_\epsilon)}
\ee
For a massless condensate, $T_{crit}\cong 5111$ GeV, and
for temperatures above top quark mass, the degeneracy factor
$g(m_t)={429 \over 4}$ is relatively
weakly dependent upon any new degrees of freedom.
To a few percent, the temperature of de-coherence is determined to be
\be
k_B T_\epsilon \: \cong \: 1592 \: GeV.
\ee
For a cold massive condensate, we will see in the subsection on Bose
condensation that the critical temperature is $T_{crit} \cong 104$ GeV, and
(assuming a degeneracy factor of $g(m_b)={345 \over 4}$)
the temperature of de-coherence is
\be
k_B T_\epsilon \: \cong \: 15.18 \: GeV.
\ee

We will next estimate the present day scale corresponding to the dark energy
de-coherence scale $R_\epsilon$.  We will assume a relatively sharp transition
from a radiation dominated expansion to a matter dominated expansion at
the dust transition red-shift, corresponding to equal energy densities of the 
(present day) relativistic and non-relativistic particles.  We will use a value
calculated from standard references\cite{PDG} for 
$z_{eq} \equiv z_{dust} \cong 3629$, where the red shift satisfies the usual
formula
\be
{\nu(z) \over \nu_o} \: \equiv \: 1+z \: = \: {R_o \over R(z)}.
\ee
After relativistic radiation falls out of equilibrium, its temperature
satisfies $T \sim 1/R$.  Photons fell out of equilibrium at last scattering $z \sim 1100$,
whereas neutrinos fell out of equilibrium much sooner at a temperature $T \sim 1 MeV$.
The present cosmic background
photon temperature is $2.725K \cong 2.35 \times 10^{-13} GeV$, and that of neutrinos is about 1.9K.  We will
calculate the red shift from CMB photon temperature to de-coherence temperature
using equation \ref{redshift}.

For a massless condensate, the red shift at decoherence is found to be
$z_\epsilon \: \approx \: 10^{16}$,
whereas for a massive condensate
$z_\epsilon \: \approx \:  10^{14}$.
We can use these redshifts to determine the present scale
associated with the de-coherence scale $R_\epsilon$.  
For a massless condensate 
$R_o \:  \approx \:  10^{14}$ cm, which is about the distance
of Satern from Earth.
For a massive condensate, this scale is given by $10^{16}$ cm, two orders
of magnitude larger.

We next examine the entropy of the system during the de-coherence period.  
The Fleisher-Susskind\cite{Susskind} entropy limit considers a black hole
as the most dense cosmological object, limiting the entropy according to
\be
S \: \leq \: S_{_{hole} ^{black}} \: = \:
{k_B c^3 \over \hbar}{A \over 4 G_N}
\ee
For a radiation dominated cosmology at de-coherence,
the entropy is proportional to the number of quanta, and is related
to the energy density $\left ( {c \over \dot{R}_\epsilon} \right )^2 \cong
{8 \pi G_N \over 3} \rho_{FL}$ by
\be
{S \over V} \: = \: {4 \over 3} {\rho_{FL} \over T_\epsilon} \Rightarrow
S \: = \: {4 \over \pi G_N} {R_\epsilon \over T_\epsilon}.
\ee
Examining this for the space-like area given by the box $A=6(2 R_\epsilon)^2$
the ratio of the entropy in a thermal environment to the limiting entropy
during thermalization is given by
\be
{S \over A/4G_N} \: \cong \: {2 \over 3 \pi} \,
\left ( {1 \over R_\epsilon T_\epsilon} \right ) 
\sim 10^{-16}.
\ee
Clearly this result satisfies the FS entropy bound regardless of the mass of the
condensate.  

\subsection{Bose condensation}
\indent

We next calculate the critical temperature for condensation of
a non-interacting gas of massless Bose quanta just prior to
dark energy decoherence.  At temperature $T$, such a gas
has energy density satisfying the relation
\be
\rho \: = \: \rho_{GS} + {\pi^2 \over 30} 
\left( {k_B T \over \hbar c} \right )^3 k_B T .
\ee
Here $\rho_{GS}$ is the density of the condensate. 
Critical temperature is defined when the second term is
insufficient to contain all particles.  For the pre-coherent
state, this is given by
\be
\rho_{FL} \: = \: {\pi^2 \over 30} {(k_B T_{crit})^4 \over (\hbar c)^3} ,
\ee
where $\rho_{FL}={3 \over 8 \pi} \left ( {M_P \over R_\epsilon} \right ) ^2 - \rho_\Lambda$ as before.
The critical temperature therefore satisfies $T_{crit}=(g(T_\epsilon))^{1 \over 4} T_\epsilon$.
This corresponds to a temperature of around 
$T_{crit}\approx 5109$ GeV for a pre-thermalized system
consisting of only gravons.  We expect the system to remain
in a zero temperature state prior to de-coherence, defining a
vacuum energy density $\rho_\Lambda$ just prior to de-coherence. 
The thermodynamics after the availability of sub-luminal degrees
of freedom will define the temperature of thermalization using
\be
\rho_{FL} \: = \: (g(T_\epsilon) +1) {\pi^2 \over 30} {(k_B T_\epsilon)^4 \over (\hbar c)^3}  +
\rho_{GS},
\ee
where $g(T_\epsilon)$ counts the degrees of freedom available to
luminal and sub-luminal interactions.  Because of the availability of the
new degrees of freedom, one expects a solution without condensate, i.e. $\rho_{GS}=0$,
to be consistent at these temperatures.

Just as de-coherence begins, we expect the fraction of condensate to
thermal gravons to satisfy
\bea
{\rho_{condensate} \over \rho_{FL}} \: = \:
1 - \left(   {T_\epsilon \over T_{crit}}   \right ) ^4 \\
{N_{condensate} \over N_{thermal}} \: = \:
1 - \left(   {T_\epsilon \over T_{crit}}   \right ) ^3 ,
\eea
where the total number of thermal gravons satisfies
\be
{N_{thermal} \over V_\epsilon} \: = \:
{\zeta (3) \over \pi^2 } 
\left ( {k_B T_{crit} \over \hbar c}  \right ) ^3 .
\ee
The de-coherent temperature is considerably lower than this
temperature due to the degrees of freedom $g(T_\epsilon)$, giving
${\rho_{condensate} \over \rho_{FL}}  \cong 0.99$ and
${N_{condensate} \over N_{thermal}} \cong 0.97$ starting thermalization.
Expressing $\rho_{FL}$ in terms of the pre-coherent condensate, we
obtain a relationship between the pre-coherent and thermal gravons,
most of which initially remain in condensate form:
\be
\rho_{FL}= {\pi^4 \over 30 \zeta (3)} {N_{thermal} \over V_\epsilon} k_B T_{crit}
=2 N_\epsilon \rho_\Lambda .
\ee
This gives a large ratio of pre-coherent to thermal gravons given by
\be
{N_\epsilon \over N_{thermal}} \: = \: {\pi^4 \over 60 \zeta (3)}
{k_B T_{crit} \over \rho_\Lambda V_\epsilon} \sim  10^{15} .
\ee
Therefore, pre-coherent gravons must rapidly thermalize a large number
of states.

We next examine the properties of a (non-interacting) Bose gas
of particle of mass $m$ for a system with temperature
$T _\sim ^ < m$. At temperature $T$, such a fluid
has energy density satisfying the relation
\be
\rho_m (T) \: = \: \rho_{GS} + 
{\zeta (3/2) \Gamma (3/2) m c^2 + \zeta (5/2) \Gamma (5/2) k_B T 
\over (2 \pi)^2 \hbar^3} 
\left( 2 m k_B T \right )^{3/2} 
\ee
where $\rho_{GS}={N_{condensate} \over V_\epsilon}
\sqrt{m^2 + k_\epsilon ^2} \cong m {N_{condensate} \over V_\epsilon}$.
Critical temperature for the pre-coherent
state is again determined when $\rho_{GS}=0$.
This corresponds to a temperature of around 
$T_{crit}\approx 103.8$ GeV for a pre-thermalized system
consisting of only scalar particles $m\cong 19.74$GeV.  
The thermodynamics after the availability of sub-luminal degrees
of freedom will define the temperature of thermalization using
\be
\rho_{FL} \: = \: g(T_\epsilon)  {\pi^2 \over 30} {(k_B T_\epsilon)^4 \over (\hbar c)^3}  +
\rho_m (T_\epsilon) .
\ee
Again, a solution without condensate $\rho_{GS}$
is consistent after thermalization, at a temperature of
de-coherence given by $T_\epsilon \cong 15.18$ GeV.

Just as de-coherence begins, we expect the fraction of condensate to
thermal scalar masses to satisfy
\be
{N_{condensate} \over N_{thermal}} \: = \:
1 - \left(   {T_\epsilon \over T_{crit}}   \right ) ^{3/2} ,
\ee
where the total number of thermal scalars satisfies
\be
{N_{thermal} \over V_\epsilon} \: = \:
{\zeta (3/2) \Gamma (3/2) \over (2 \pi)^2 \hbar^3} 
\left( 2 m k_B T_{crit} \right )^{3/2}  .
\ee
As decoherence begins, the condensate density fraction is given by
${\rho_{condensate} \over \rho_{FL}}  \cong 0.98$ and
${N_{condensate} \over N_{thermal}} \cong 0.94$ starting thermalization.
After de-coherence, ${\rho_m \over \rho_{FL}} \cong 0.018$, which
means that less than 2 \% of the thermalized matter
is made up of masses $m$.  
The relationship between the pre-coherent and thermal scalars is then
given by
\be
\rho_{FL}= \left ( m +
{\zeta (5/2) \Gamma (5/2) \over \zeta (3/2) \Gamma (3/2)} k_B T_{crit} \right )
{N_{thermal} \over V_\epsilon} 
=2 N_\epsilon \rho_\Lambda .
\ee
This gives a ratio of pre-coherent to thermal masses $m$ given by
\be
{N_\epsilon \over N_{thermal}}  \sim 5.
\ee
This means that the thermalization process for a massive condensate
is not as severe as that for a massless condensate.

\subsection{Pre-coherence}
\indent

Although up to this point we have avoided examining the cosmology
prior to dark energy de-coherence, it is useful to conjecture on the
continuity of the physics of this period.  Since the cosmological
scale expansion is supraluminal, only gravitational interactions
are available for cosmological equilibrations.  We will assume
that the cosmological scale excitations will have energies that
satisfy the usual Planck relation, only with propagation speed
determined by the expansion rate:
\be
E_\epsilon \, = \, h \nu \, = \, {h \dot{R} \over \lambda}
\, = \, {h \dot{R} \over 2 R} \, = \, \pi \hbar H
\ee
For the scalar long range gravitating quanta (collective modes)
discussed previously, the density of states is expected to
be of the form
\be
\Delta^3 n \: = \: {V \over (2 \pi)^3} d^3 k \: = \: 
{4 \pi \over (\pi \hbar)^3 }{E^2 d E \over H^3}
\ee

\textit{If} there is thermal equilibration, we therefore expect the usual forms for a scalar
boson, with the substitution $\hbar c \rightarrow \hbar \dot{R}$.  In particular,
the energy density takes the form
\be
\rho \: = \: {\pi^2 \over 30} {(k_B T)^4 \over (\hbar \dot{R})^3} \: = \:
{\pi^2 \over 30} \left ( {1 \over \hbar R} \right )^3 {(k_B T)^4 \over H^3}.
\ee
We assume that the FL equation continues to drive the dynamics, which
allows substitution of the Hubble rate in terms of density
\be
\rho \: = \: {\pi^2 \over 30} \left ( {1 \over \hbar R} \right )^3 
{(k_B T)^4 \over \left [ {8 \pi G_N \over 3} (\rho + \rho_\Lambda) \right ] ^{3 \over 2}}.
\ee
Since these gravons are expected to behave like radiation 
${\rho \over \rho_{FL}}=\left( {R_\epsilon \over R} \right )^4$
(as any condensate is likewise expected to consistently scale), we determine
the scaling of temperature with cosmological scale
\be
\left ( {R_\epsilon \over R} \right ) ^7 \: = \: \left ( {T \over T_\epsilon} \right ) ^4
\ee
Thus, we see that the scaling of temperature with inverse FRW scale factor
no longer holds.  As suspected, the equation of state is considerably altered
prior to dark energy de-coherence.

If we consistently continue this conjecture to determine the critical temperature
for Bose condensation of the gravons, the number of quanta in a scale volume
is given by
\be
N \: = \: N_{condensate} \, + \, {\zeta (3) \over \pi^2} \left ( {T \over \hbar H} \right ) ^3 .
\ee
As usual, the ratio of condensate to "normal" state satisfies
\be
{N_{condensate} \over N } \: = \: 1 - \left ( {T \over T_c} \right ) ^3 .
\ee
The critical temperature is given by
\be
k_B T_c \: = \: \left ( {\pi ^ 2 \over 8 \zeta (3)} \right ) ^{1/3} \hbar
\sqrt{{8 \pi G_N \over 3} \rho }.
\ee
Therefore, since the energy density is expected to scale like $R^{-4}$, we
can conclude that the critical temperature scales as
\be
{T_c \over T_{c \epsilon}} \: = \: \left ( {R_\epsilon \over R} \right )^2
\: = \: \left ( {T \over T_\epsilon} \right ) ^{{8 \over 7}}.
\ee
Since $T_c$ increases more rapidly than $T$ at higher temperatures,
such a system would remain condensed at early times.  This means
that a system obeying this behavior would have a suppressed vacuum
energy due to the condensation into the lowest momentum mode until
thermalization during de-coherence.  For such a system, the coherence
of a supraluminal horizon  need not be driven by the rapid
expansion rates, but rather is a direct consequence of the global
quantum coherence of the macroscopic quantum system.

Since the ratio of the temperature to the critical temperature
becomes vanishingly small for the earliest times
\be
{T \over T_c} \sim 0.31 \left ( {T_\epsilon \over T}  \right )^{1/7} \Rightarrow 0
\ee
we feel justified in asserting that the pre-coherent cosmology
which starts completely condensed will remain
a zero temperature condensate until de-coherence. 
This condition is required to justify the use of the lowest momentum
mode only in the evaluation of cosmological vacuum energy density
at de-coherence.

\subsection{Fluctuations}
\indent

Adiabatic perturbations are those that fractionally perturb the number
densities of photons and matter equally.  For adiabatic perturbations,
the energy density fluctuations grow according to\cite{PDG}
\be
\delta \: = \: \left \{
\begin{array}{cc}
\delta_\epsilon  \left ( {R(t) \over R_\epsilon} \right ) ^2  & 
radiation-dominated \\
\delta_{dust} \left ( {R(t) \over R_{dust}} \right ) &
matter-dominated .
\end{array}
\right .
\ee
Temperature fluctuations are expected to be related to density
fluctuations using ${\delta T \over T} \cong {1 \over 3} {\delta \rho \over \rho}=
{1 \over 3} \delta$.
This allows us to write an accurate estimation for the scale of fluctuations
during de-coherence in terms of those at last scattering
\be
\delta_\epsilon \: = \: \left( {R_{dust} \over R_{LS}}  \right)
\left( {R_\epsilon \over R_{dust}}  \right) ^2 \delta_{LS} \: \cong \:
{z_{dust} z_{LS} \over z_\epsilon ^2} \delta_{LS} 
\ee
if the fluctuations are ``fixed" at dark energy de-coherence.  Assuming the
values for $z_{dust}$ and $z_\epsilon$ calculated previously, along with
the red shift at last scattering $z_{LS}\approx 1100$, this
requires fluctuations fixed at dark energy de-coherence to have a value
\bea
\delta_\epsilon \: \approx \: 8.46 \times 10^{-27} \delta_{LS} \: \sim \: 10^{-31} \quad massless \\
\delta_\epsilon \: \approx \: 1.63 \times 10^{-22} \delta_{LS} \: \sim \: 10^{-27} \quad massive.
\eea
If quantum coherence persists such that the
fluctuations are fixed at a later scale $R_F$, this relation gets modified
to take the form
\be
\delta_F \: \cong \: {z_{dust} z_{LS} \over z_F ^2} \delta_{LS} \: \approx \:
\delta_\epsilon \left ( {R_F \over R_\epsilon} \right )^2 .
\ee

We expect the energy available for fluctuations to be of the order of
the vacuum energy.  This energy drives the two-point correlation
function for the squared deviations from the average density,
which means that we should expect the amplitude of
the fluctuations to be of the order
\be
\delta_{DC} \: \sim \: \left (  {\rho_\Lambda \over \rho_{FL} + \rho_\Lambda }   \right ) ^{1/2} 
\: \cong \: \left (  {1 \over 2 N_\epsilon} \right ) ^{1/2},
\ee
regardless of the specifics of the condensate.
This form also appears in the literature on fluctuations\cite{Harrison}.
Indeed, we obtain the correct order of magnitude for fluctuations at de-coherence
for either massless or massive condensates
\be
\delta_{DC} \: \cong \: \left \{
\begin{array}{l}
3.7 \times 10^{-31}  \quad massless \\
4.5 \times 10^{-27} \quad massive.
\end{array}
\right .
\ee
At last scattering this gives
\be
\delta_{LS} \: \cong \: \left \{
\begin{array}{l}
3.0 \times 10^{-4}  \quad massless \\
2.8 \times 10^{-5} \quad massive,
\end{array}
\right .
\ee
whereas for present day observations, this fluctuation is given by
\be
\delta_o \: \cong \: \left \{
\begin{array}{l}
0.3  \quad massless \\
0.03 \quad massive,
\end{array}
\right .
\ee
if the fluctuation grows only linearly (which is not the case for late times).
The amplitude of galaxy fluctuations is expected to be $\sigma_8 \cong 0.84$,
which is the linear prediction theoretical prediction for the amplitude of
fluctuations within 8 Mpc/h spheres\cite{Spergel}.
We see that the massive condensate best matches fluctuations at last
scattering (i.e. $\sim 10^{-5}$), but exploration of the agreement with present day fluctuations
requires more than our simple extrapolation from last scattering.

\section{Discussion and Conclusions}
\indent

We feel that we have given a strong argument for the interpretation
of cosmological dark energy as the vacuum energy of a zero
temperature condensate of bosons.
Prior to de-coherence, the scale of gravitational 
quantum vacuum energy is given by the Friedman-Robertson-Walker (FRW) scale $R(t)$.  
We have asserted that dark energy de-coherence occurs when
$\dot{R}=c$, which is only consistent with a spatially flat cosmology. 
During de-coherence, the gravitational coherence scale of the Friedman-Lemaitre (FL)
density changes considerably  
(most likely to be the Compton wavelength of the mass $m$ associated with the
Bose condensate, which is much less than the coherence scale 
of the dark energy), resulting in a gravitational phase transition, and the onset of new thermal
degrees of freedom.  
This means that microscopic thermal interactions
between components of the FL energy will break gravitational  coherence, freezing the value of
the gravitational dark energy.  We have assumed that the quantum 
vacuum state for gravitation is an intrinsic state, with an energy density scale given by 
the vacuum energy density of the zero temperature Bose condensate
during the period of last quantum coherence 
(given by $\epsilon ={1 \over 2} \sqrt{m^2 + k_\epsilon ^2}$, which is determined by the
current value of the cosmological constant).  When the cosmology has global coherence, 
the gravitational vacuum state is expected to evolve with the contents of the universe.  
When global coherence is lost, there remains only local coherence 
within independent clusters, and the prior vacuum state loses scale coherence 
with the clusters as the new degrees of freedom become available.    This dark energy scale 
will be frozen out as a cosmological constant of positive energy density satisfying 
$\rho_\Lambda \: = \: {\Lambda \over 8 \pi G_N}= {\epsilon k_\epsilon ^3 \over (2 \pi)^3 }$
in terms of the present day cosmological constant.

To determine the ultimate fate of the universe, one needs an understanding
of the fundamental nature of the quantum vacuum.  The Wheeler-Feynman
interpretation of the propagation of quanta irreducibly binds those quanta to
their sources and sinks.  A previous paper by these authors directly
demonstrates the equivalence of the usual Compton scattering process
calculated using photons as the asymptotic states in standard QED with
a description that explicitly includes the source and sink of the scattered
photons in a relativistic three-particle formalism\cite{LN2}.
According to Lifshitz and others\cite{Lifshitz}, the zero
temperature electromagnetic field in the Casimir effect can be derived in
terms of the zero-point motions of the sources and sinks upon which the
forces act.  In the absence of a causal connection between those sources
and sinks, one has a difficult time giving physical meaning to a
vacuum energy or Casimir effect.  Since the zero-point motions produce
classical electromagnetic fields in Landau's treatment, these fields
propagate through the``vacuum" at c. This would mean that
one expects the Casimir effect to
be absent between comoving mirrors in a cosmology with $\dot{R}>c$. 
If the regions in a future cosmology whose expansion are driven by vacuum energy
are indeed causally disjoint, then there could be no driving of that expansion
due to the local cosmological constant.  Such an expansion requires that
gravitational interactions propagate in a manner that causally affects
regions requiring super-luminal correlations.  The expected change in
the equation of state for the cosmology as a whole should modify
the behavior of the FRW expansion in a manner that would require
reinterpretation of the vacuum energy term, as seems to be necessary during
the pre-coherence epoch.

This means that we do not view the cosmological constant as the \textit{same} 
as vacuum energy density.  In our interpretation, cosmological vacuum energy changes during pre-coherence
and post-coherence.  The cosmological constant is frozen at de-coherence due to the
availability of luminal degrees of freedom.  Since this vacuum energy density is associated 
with regions of global gravitational coherence, it is interesting to
consider whether subsequent expansion in space-time will re-establish
coherence on a cosmological scale. 

We are in the process of examining the power spectrum of fluctuations expected to
be generated by de-coherence as developed.  It is our hope that further explorations
of the specifics of density and temperature fluctuations will allow us to better differentiate
between massless vs various massive condensates. 
In addition, we have begun to examine
the scale of the symmetry breaking involved in coherence, especially with regards
to the mass scales involved.  It is our belief that this approach will reduce the
parameter set needed to describe a consistent cosmology.

\begin{center} {\bf Acknowledgment}
\end{center}

We are most grateful, once again, to E.D.Jones for permitting us
to discuss and extend his theory before his own paper is
available.

\end{document}